\documentstyle[epsf,psfig]{mn}

\newif\ifAMStwofonts

\ifoldfss
  \ifCUPmtlplainloaded \else
    \NewTextAlphabet{textbfit} {cmbxti10} {}
    \NewTextAlphabet{textbfss} {cmssbx10} {}
    \NewMathAlphabet{mathbfit} {cmbxti10} {} 
    \NewMathAlphabet{mathbfss} {cmssbx10} {} 
  \fi
  \ifAMStwofonts
    \ifCUPmtlplainloaded \else
      \NewSymbolFont{upmath} {eurm10}
      \NewSymbolFont{AMSa} {msam10}
      \NewMathSymbol{\upi}     {0}{upmath}{19}
      \NewMathSymbol{\umu}     {0}{upmath}{16}
      \NewMathSymbol{\upartial}{0}{upmath}{40}
      \NewMathSymbol{\leqslant}{3}{AMSa}{36}
      \NewMathSymbol{\geqslant}{3}{AMSa}{3E}

    \fi
  \fi
\fi 

\ifnfssone
  \newmathalphabet{\mathit}
  \addtoversion{normal}{\mathit}{cmr}{m}{it}
  \addtoversion{bold}{\mathit}{cmr}{bx}{it}
  \newmathalphabet{\mathbfit} 
  \addtoversion{normal}{\mathbfit}{cmr}{bx}{it}
  \addtoversion{bold}{\mathbfit}{cmr}{bx}{it}
  \newmathalphabet{\mathbfss} 
  \addtoversion{normal}{\mathbfss}{cmss}{bx}{n}
  \addtoversion{bold}{\mathbfss}{cmss}{bx}{n}
  \ifAMStwofonts
    \ifCUPmtlplainloaded \else
      \UseAMStwoboldmath
      \makeatletter
      \new@mathgroup\upmath@group
      \define@mathgroup\mv@normal\upmath@group{eur}{m}{n}
      \define@mathgroup\mv@bold\upmath@group{eur}{b}{n}
      \edef\UPM{\hexnumber\upmath@group}
      \new@mathgroup\amsa@group
      \define@mathgroup\mv@normal\amsa@group{msa}{m}{n}
      \define@mathgroup\mv@bold\amsa@group{msa}{m}{n}
      \edef\AMSa{\hexnumber\amsa@group}
      \makeatother
      \mathchardef\upi="0\UPM19
      \mathchardef\umu="0\UPM16
      \mathchardef\upartial="0\UPM40
      \mathchardef\leqslant="3\AMSa36
      \mathchardef\geqslant="3\AMSa3E
    \fi
  \fi
\fi 

\ifnfsstwo
  \DeclareMathAlphabet{\mathbfit}{OT1}{cmr}{bx}{it}
  \SetMathAlphabet\mathbfit{bold}{OT1}{cmr}{bx}{it}
  \DeclareMathAlphabet{\mathbfss}{OT1}{cmss}{bx}{n}
  \SetMathAlphabet\mathbfss{bold}{OT1}{cmss}{bx}{n}
  \ifAMStwofonts
    \ifCUPmtlplainloaded \else
      \DeclareSymbolFont{UPM}{U}{eur}{m}{n}
      \SetSymbolFont{UPM}{bold}{U}{eur}{b}{n}
      \DeclareSymbolFont{AMSa}{U}{msa}{m}{n}
      \DeclareMathSymbol{\upi}{0}{UPM}{"19}
      \DeclareMathSymbol{\umu}{0}{UPM}{"16}
      \DeclareMathSymbol{\upartial}{0}{UPM}{"40}
      \DeclareMathSymbol{\leqslant}{3}{AMSa}{"36}
      \DeclareMathSymbol{\geqslant}{3}{AMSa}{"3E}
    \fi
  \fi
\fi 

\ifCUPmtlplainloaded \else
  \ifAMStwofonts \else 
    \def\upi{\pi}
    \def\umu{\mu}
    \def\upartial{\partial}
  \fi
\fi

\title{Disc loss and renewal in A0535+26}

\author[N.J.Haigh, M.J.Coe, I.A.Steele and J.Fabregat]
       {N.J.Haigh$^{1}$, M.J.Coe$^{1}$, I.A.Steele$^{2}$ and J.Fabregat$^{3}$ \\
$^{1}$Department of Physics and Astronomy, The University, Southampton, SO17 1BJ, UK.\\ \\
$^{2}$Astrophysics Reseearch Institute, Liverpool John Moores University, Twelve Quays House, Egerton Wharf, Birkenhead, CH41 1LD, UK.\\ \\
$^{3}$Departamento de Astronomia y Astrofisica, Universidad de Valencia, E-46100 Burjassot, Valencia, Spain. \\ \\}

\date{Accepted 1999 October 6. Received 1999 August 2}

\pagerange{\pageref{firstpage}--\pageref{lastpage}}
\pubyear{1999}

\begin{document}

\maketitle

\label{firstpage}

\begin{abstract}
This paper presents observations of the Be/X-ray binary system A0535+26 revealing the first observed loss of its circumstellar disc, demonstrated by the loss of its JHK infrared excess and optical/IR line emission. However optical/IR spectroscopy reveals the formation of a new inner disc with significant density and emission strength at small radii; the disc has proven to be stable over 5 months in this intermediate state. 

\end{abstract}

\section{Introduction}

HDE 245770 is the 09.7-B0IIIe (Steele et al. 1998) optical counterpart of the classical Be/X-ray binary system A0535+26, discovered in 1975 (Coe et al. 1975; Rosenberg et al. 1975). The system contains a 104s spin-period neutron star in a 111 day orbit (Hayakawa 1981).

Optical/IR spectra have revealed H$\alpha$, Br$\gamma$, Pa$\gamma$ and Pa$\delta$ to be in emission throughout the observed history of the system, suggesting the continuous presence of a circumstellar disc. However, observations reported in this paper reveal a period of absorption in the above H lines, though H$\alpha$ has rapidly resumed very weak emission. Combined with a major decrease in IR luminosity, the implication is that a large reduction in the circumstellar disc characteristic of this Be/X-ray binary must have occurred. Further IR/optical spectra presented here reveal the rapid re-growth and ensuing stability of a small, dense inner disc.

\begin{table*}
\caption{Observations.}
\begin{tabular}{llll}

Date              & Telescope & Instrument & Details\\
&&&\\
1987-1998         &TCS                &CVF                       &JHK photometry\\
30 August 1998    &KPNO Coud\'{e}-Feed&Coud\'{e} CCD Spectrograph&F3KB, 108cm camera, 316 l/mm (Grating B)\\
10 September 1998 &KPNO Coud\'{e}-Feed&Coud\'{e} CCD Spectrograph&F3KB, 108cm camera, 316 l/mm (Grating B)\\
10 November 1998& UKIRT               & CGS4& ZJHK spectra, long camera, 256$^2$ array, 40 l/mm grating \\
26 November 1998& INT                 & IDS & TEK5, 235 camera, R600R grating\\
15 December 1998& JKT                 & JAG-CCD camera           & UBVRI photometry\\
24 December 1998& WHT                 & ISIS& TEK2, R600R grating\\
6 January 1999  & SAAO 1.9m           & Mk III Photometer        & JHK photometry\\
11 January 1999 & SAAO 1.9m           & Spectrograph             & SITe, 1200l/mm grating\\
23 January 1999 & SAAO 1.0m           & CCD camera               & UBVRI photometry\\
4 February 1999 & INT                 & IDS & TEK5, 235 camera, R1200R grating\\
22 February 1999& INT                 & IDS & TEK5, 235 camera, R1200R grating\\
4 March 1999    & INT                 & IDS & 500m camera, 1200l/mm grating\\
7 March 1999    & INT                 & IDS & TEK5, 235 camera, R1200R grating\\
24 April 1999   & INT                 & IDS & TEK5, 235 camera, R1200Y grating\\

\label{tab:obs}
\end{tabular}
\end{table*}

\section{Observations}

A summary of all the observations in this work may be found in Table~\ref{tab:obs}. All data have been reduced using Starlink software, with the exception of the August and September 1998 spectra (which were reduced within IRAF), SAAO IR photometry and TCS photometry. The latter were obtained from the 1.5m Telescopio Carlos Sanchez at Teide observatory, Tenerife using the CVF photometer. Data were reduced by means of the procedure described by Manfroid (1993). Instrumental values were transformed to the TCS standard system (Alonso et al. 1998). SAAO IR photometry was reduced using standard SAAO software which is also based upon the methods described in Manfroid (1993).

Optical spectra have mostly been obtained via the service observing programme of the ING, and all include H$\alpha$ and He {\sc i} 6678$\AA$. Raw spectra were de-biased, flat-fielded, extracted and wavelength calibrated (using CuAr and ArNe arc lamp spectra) with the {\footnotesize FIGARO} package. Heliocentric velocity corrections were applied using {\footnotesize VACHEL}. Further processing made use of {\footnotesize DIPSO}. Image reduction was undertaken using the {\footnotesize FIGARO} package, whilst subsequent aperture photometry took place within {\footnotesize GAIA}.

Infrared spectra of the system were obtained on 1998 November 10 using the CGS4 spectrometer of the United Kingdom Infrared Telescope (UKIRT), Hawaii. Four grating positions were employed (0.95 -- 1.26, 1.19 -- 1.34,  1.40 -- 1.80, and 1.70 -- 2.40 $\mu$m), giving almost complete spectral coverage from 0.95 to 2.4 $\mu$m. A and a G-type standards were also observed in order to allow correction of the spectra for telluric absorption.

Initial data reduction was carried out at the telescope using the {\footnotesize CGS4DR} software (Puxley, Beard \& Ramsey 1992). Subsequent sky subtraction, optimal extraction, division by the standard star, and wavelength calibration using arc lamp observations were carried out using {\footnotesize FIGARO}.

\section{Discussion}

\subsection{IR and Optical photometry}
 
Photometric variability in Be/X-ray binaries is largely attributable to changes in the disc, particularly towards the JHK bands where disc flux can be dominant during optically bright states; Roche et al. (1993) noted cessation of all optical variability during periods of disc loss ('faint states') in the system X Per. Discussion of the previous photometric, spectroscopic and X-ray behaviour of A0535+26 may be found in Clark et al. (1998a, 1998b).

\begin{table}
\caption{Recent optical/IR photometry and March/April 1991 data for comparison (Clark (1997) and this work.) CRAO BV data are from 10 March 1991, RI data are from 5 April 1991. 1991 TCS data are from 13 April.}
\begin{tabular}{|c|c|c|c|c|c|}
  &        &           &      & 1991  &1991\\
  & JKT    & TCS       & SAAO & CRAO  &TCS\\
  &        &           &      &       &\\
B & 9.85   &           & 9.92 & 9.63  &\\
V & 9.44   &           & 9.46 & 9.16  &\\
R & 9.12   &           & 9.13 & 8.36  &\\
I & 8.83   &           & 8.80 & 7.95  &\\
J &        & 8.49      & 8.62 &       &7.72\\
H &        & 8.39      & 8.42 &       &7.48\\
K &        & 8.34      & 8.34 &       &7.28\\

\label{tab:photom}
\end{tabular}
\end{table}

\begin{figure}
\begin{center}
\psfig{file=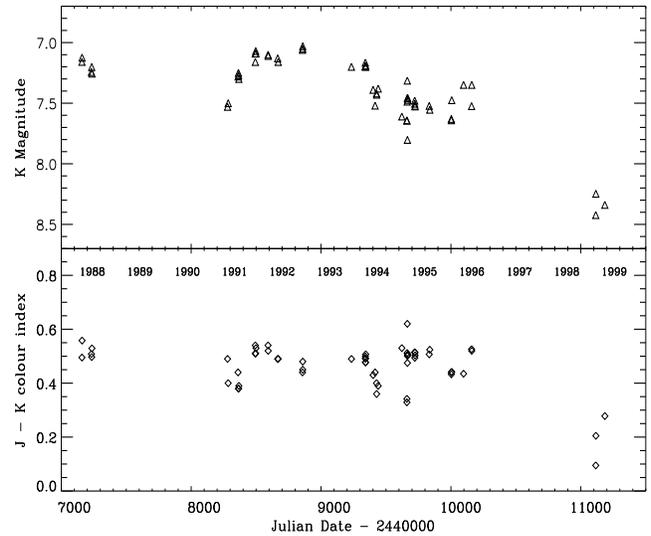,width=8.4cm,height=7cm,angle=90}
\caption{The last 12 years of K band magnitudes and J-K colour indices.}
\label{fig:jhk}
\end{center}
\end{figure}

Comparison of October 1998 JHK photometry with archive data (this work and Clark et al. 1998b) showed that a marked fading in all three wavebands had occurred since the previous observation in March 1996, the decrease in K being 0.8 magnitudes (Figure~\ref{fig:jhk}, top). The system has also become bluer, the J-K colour index decreasing by approximately 0.3 (Figure~\ref{fig:jhk}, bottom).

A good example of a typical system state can be seen in March/April 1991 (chosen because there exists quasi-simultaneous UBVRIJHK photometry and spectra). At this time the disc was in a characteristically strong emission state with J and K magnitudes of 7.72 and 7.28 respectively (Table~\ref{tab:photom}). On 27-28 October 1998 these magnitudes were 8.49 and 8.34, i.e., the fluxes in these bands had fallen to $\sim$50$\%$ and $\sim$40$\%$ of their April 1991 values. These percentages are close to the star/(star+disc) emission fraction as determined from Roche et al. (1993) for X Per, deduced after this similar system suffered a disc loss event in 1990 which saw its IR fluxes fall in a similar manner. The lower, bluer colour index reflects the greater relative contribution of the hot B star flux. These results demonstrate the loss of the bulk of the infrared emission from the circumstellar disc around HDE~245770.

Optical magnitudes (BVRI) from both recent data sets are significantly fainter than seen historically, consistent with the loss of the disc contribution, and the commencement of a `faint state'.

\subsection{Optical Spectroscopy}

\begin{figure}

\psfig{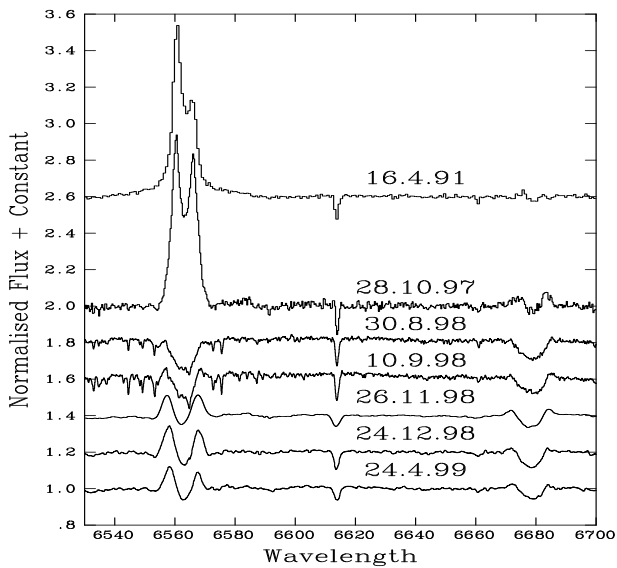}
\caption{Representative spectra showing recent changes. The `typical' March/April 1991 state is shown at the top. Below is the final spectrum before disc-loss. The lower five spectra show the re-forming disc and subsequent standstill. The 1991 and 1997 spectra are from Clark et al. (1998a).}
\label{fig:halphahist}

\end{figure}

The KPNO spectrum of 30 August 1998 (Gies, private communication) reveals almost complete loss of emission at H$\alpha$, a sure sign of very significant disc loss (Figure~\ref{fig:halphahist}). Comparison with older archive spectra from Clark et al. (1998a) shows that dramatic changes in spectral profile have occurred. Prior red spectra feature H$\alpha$ emission (single or double peaked) with a large equivalent width; in the case of spectra (Figure~\ref{fig:halphahist}) from April 1991 the EW is -7.5$\AA$.

By September 10 symmetrical emission wings had formed in H$\alpha$, and the R peak of He {\sc i} 6678$\AA$ had strengthened. Spectra appeared to reach a stable state by the 26 November when the first INT IDS service spectrum was obtained.

\begin{figure}
\begin{center}
\psfig{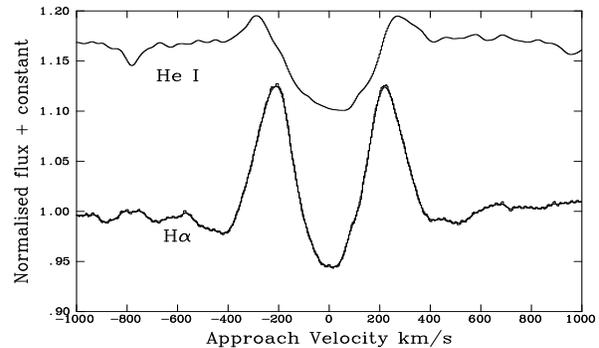}
\caption{H$\alpha$ and He {\sc i} 6678$\AA$ profiles from the mean of all spectra obtained during the stable state (26 November 1998 to 24 April 1999).  Note the differing peak rotational velocities.}
\label{fig:velplot}
\end{center}
\end{figure}

All spectra from 26 November onwards (most recent being 24 April) display weak double peaked emission either side of photospheric absorption below the continuum (best seen in Figure~\ref{fig:velplot}). The EW remains at approximately -0.5$\AA$, barely in emission. The period during which the disc loss occurred is constrained by optical spectra to lie between 28 October 1997 and 30 August 1998.

In order to monitor the expected rebuilding of the disc, spectra have been acquired as frequently as possible; most spectra are from INT and WHT service requests. Previous Be star disc losses have been followed by rebuilding over a timescale from weeks to years (e.g. X Per, Roche et al. 1993). However spectra from seven further epochs, spanning 5 months, have shown no major changes, though small variations in line profile have certainly occured. Thus the situation with A0535+26 is all the more interesting because it represents a stable state intermediate between normal and complete disc loss.

Assuming that displacement of line profile features from rest wavelength is largely a kinematic effect (Huang 1972), information regarding distribution of emitting gas around the star can be inferred from the line profile. High velocity emission wings are the signature of a small, rapidly rotating inner disc; the absorption core of H$\alpha$ (Figure~\ref{fig:velplot}) is consistent with the assertion from IR photometry that most of the disc has lost much of its emission and presumably mass; it reveals a strongly depleted disc outwards of several stellar radii. The data are consistent with the idea that H$\alpha$ is emitted from similar regions in the disc as the JHK excess.

Figure~\ref{fig:velplot} shows the peaks of He {\sc i} 6678$\AA$ emission at 280$\pm$20 km$s^{-1}$ and at 215$\pm$20km$s^{-1}$ for H$\alpha$. This demonstrates that hotter, denser conditions, preferred for He {\sc i} 6678$\AA$ emission, exist at smaller radii than those primarily responsible for H$\alpha$ emission (Clark et al. 1998b). Though weaker than in pre-disc-loss measurements, He {\sc i} 6678$\AA$ emission wings are present at a greater fraction of their historical intensity than H$\alpha$. This suggests that the very innermost disc has a higher fraction of its pre-loss density than the regions slightly further out giving rise to the H$\alpha$ peaks. 

Several authors have derived masses for the primary based on luminosity, spectral class and orbital motion considerations: 9-17M$_{\sun}$ Janot-Pacheco, Motch \& Mouchet (1987), 10-20M$_{\sun}$ Clark et al. (1998a), 8-22M$_{\sun}$ Wang \& Gies (1998). Assuming a mass of 15M$_{\sun}$ and an inclination of 26-40$^{\circ}$ (Wang \& Gies 1998), the 280$\pm$20 km$s^{-1}$ peak velocity of He {\sc i} 6678$\AA$ emission corresponds to an orbital radius of 7.6$\pm2.5\times$10$^{6}$km  whilst H$\alpha$ peaks yield 13$\pm5\times$10$^{6}$km. An O9.7III star is expected to have a radius of approximately $10\times$10$^{6}$km, thus the true orbital radii must lie at the upper end of this range, suggesting that the rotational inclination lies at the upper end (closer to 40$^{\circ}$) of the assumed range. In any case, the reforming disc can be seen to be restricted to very near the stellar surface. ( The `typical' April 1991 spectrum (Figure~\ref{fig:halphahist}) exhibits H$\alpha$ peaks at velocities of 125$\pm$20 km$s^{-1}$, giving a radius of 38$\pm15\times$10$^{6}$km.)

\begin{figure}
\begin{center}
\psfig{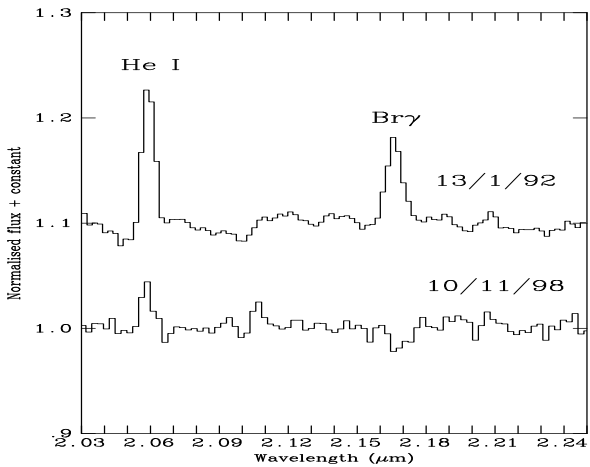}
\caption{The $K$ band spectrum obtained on 10 November 1998. For comparison spectra 
from 1992 and 1994 (Clark et al. 1998b) are also shown.}
\label{fig:kplot}
\end{center}
\end{figure}

\subsection{Infrared Spectra}

Previous IR spectra of A0535+26 have been presented by Clark et al. (1998b) and Clark et al. (1999b). Figure~\ref{fig:kplot} presents 2.03 - 2.25$\mu$m excerpts from a typical $K$ band historical spectrum and the `disc loss' spectrum from 1998 November 10.  The Br$\gamma$ 2.16 $\mu$m feature can be seen to have gone into absorption (EW=2.3 {\AA}), although the He {\sc i}  2.058 $\mu$m feature is still weakly in emission (EW=-2.8 {\AA}).  Similarly in the 0.96 - 1.14$\mu$m $Z$ band spectrum He {\sc i} 1.083$\mu$m is in emission (EW=-2.1 {\AA}), but Pa$\gamma$ 1.094$\mu$m and Pa$\delta$ 1.005$\mu$m are in absorption (EWs 1.9{\AA} and 1.7{\AA} respectively). Unlike the previous spectra (Clark et al. 1998b), the 1.4 -- 1.8$\mu$m spectrum is essentially featureless, showing no evidence for Balmer series emission.

Overall these spectra, showing He {\sc i} emission but H {\sc i} absorption are unique in our experience of IR spectra of Be stars. Clark \& Steele (1999) present $K$ band spectra of a sample of 66 Be stars; in {\em all} of those spectra whenever He {\sc i} 2.058$\mu$m emission is present it is accompanied by Br$\gamma$ emission.  It is clear therefore that we are viewing A0535+26 in a highly unusual state.

We now consider the population mechanisms for the He {\sc i} lines observed in emission. The He {\sc i} 1.083$\mu$m triplet (1s2s -- 1s2p, $^3$S -- $^3$P$^0$) can be populated either by recombination from He {\sc ii}, or collisionally from He {\sc i}. He {\sc i} 2.058$\mu$m (1s2s -- 1s2p, $^1$S -- $^1$P$^0$) is the singlet equivalent of the 1.083$\mu$m line and is primarily populated by recombination (Clark et al. 1999b).  However it is also necessary that the environment is dense, in order that the optical depth at 584{\AA} is large enough to prevent de-excitation of the upper 1s2p state via the more favoured (1s$^2$ -- 1s2p, $^1$S - $^1$P$^0$) 584{\AA} transition.  

It can therefore be seen that the He {\sc i} 2.058$\mu$m emission indicates that a hot (to provide sufficient He {\sc ii} to populate the upper level), dense (to provide sufficient He {\sc i} 584 {\AA} optical depth) environment is required.  This implies that disc reformation was already taking place by early November 1998. The lack of H emission is most easily explained by the temperature in this region being so high (greater than around 15,000 K) that the hydrogen is almost fully ionized.

\section{Conclusions}

The Be/X-ray binary system A0535+26 has undergone a phase change involving the loss of much of its disc, as revealed by the loss of its JHK infrared excess and the bulk of its optical/IR line emission. However optical line profiles and IR line strengths have revealed the formation of a dense inner disc. The standstill in this state is noteworthy, and further detailed observations are encouraged.

\section*{Acknowledgments}

We are tremendously grateful to Don Pollacco and the ING Service Programme, Douglas Gies and Pierre Maxted for acquiring valuable spectra for us. Many thanks to Simon Clark and Ignacio Negueruela for helpful discussions and suggestions. The INT and WHT are operated on the island of La Palma by the Isaac Newton Group in the Spanish Observatorio del Roque de los Muchachos of the Instituto de Astrofisica de Canarias. NJH is in receipt of a PPARC studentship.

\bsp

\label{lastpage}

\end{document}